\documentclass[preprint,aps,showpacs,showkeys]{revtex4}
\usepackage{epsfig}
\newcommand {\be}{\begin{equation}}
\newcommand {\ee}{\end{equation}}
\newcommand {\bea}{\begin{eqnarray}}
\newcommand {\ea}{\end{eqnarray*}}
\newcommand {\ba}{\begin{eqnarray*}}
\newcommand {\eea}{\end{eqnarray}}

\newcommand {\sigij}{{\bf \sigma}_i\cdot {\bf \sigma}_j}

\newcommand {\tauij}{{\bf \tau}_i\cdot {\bf \tau}_j} 
\newcommand {\Sij}{S_{ij}}
\newcommand {\LS}{{\bf L} \cdot {\bf S}}
\newcommand {\Gw}{G(\omega )}

\newcommand{\bem}{\begin{multline}}

\newcommand{\bit}{\begin{itemize}}
\newcommand{\eit}{\end{itemize}}

\newcommand{\ic}{\begin{center}}
\newcommand{\fc}{\end{center}}
\begin{document}
\title{ Spin--orbit and tensor interactions in homogeneous 
 matter of nucleons: accuracy of modern many--body theories}
\author{I. Bombaci$^1$, A. Fabrocini$^{1,a}$, A. Polls$^2$, 
and I. Vida\~na$^{3}$}
\affiliation{$^1$Dipartimento di Fisica "E.Fermi", Universit\`a di Pisa, and  
INFN, Sezione di Pisa, Largo Bruno Pontecorvo 3, I-56127 Pisa, Italy}
\affiliation{ $^2$ Departament d'Estructura i Constituents de la
  Mat\`eria,  Diagonal 645, Universitat de Barcelona, 
  E-08028 Barcelona, Spain}
\affiliation{$^3$ Gesellschaft f\"{u}r Schwerionenforschung (GSI), 
Planckstrasse 1, D-64291 Darmstadt, Germany}
\affiliation{$^a$ Corresponding Author: 
+39-050-2214 903; e--mail: adelchi.fabrocini@df.unipi.it}

\date{\today}

\begin{abstract}
We study the energy per particle of symmetric nuclear matter and pure 
neutron matter using realistic nucleon--nucleon potentials having 
non central tensor and spin--orbit components, up to three times the 
empirical nuclear matter saturation density, $\rho_0=0.16$ fm$^{-3}$. 
The calculations are carried out within the frameworks of the 
Brueckner--Bethe--Goldstone (BBG) and Correlated Basis Functions 
(CBF) formalisms, in order to ascertain the accuracy of the methods. 
The two hole--line approximation, with the continuous choice for the 
single particle auxiliary potential, is adopted for the BBG approach, 
whereas the 
variational Fermi Hypernetted Chain/Single Operator Chain theory,  
corrected at the second order perturbative expansion level, is used 
in the CBF one. The energies are then compared with the available Quantum 
and Variational Monte Carlo results in neutron matter and with the 
BBG, up to the three hole--line diagrams. 
For neutron matter and potentials without spin--orbit components 
all methods, but perturbative CBF, are in reasonable agreement up 
to $\rho\sim$ 3 $\rho_0$. 
After the inclusion of the LS interactions, we still find agreement 
around $\rho_0$, whereas it is spoiled at larger densities. 
The spin--orbit potential lowers the energy of neutron matter at 
$\rho_0$ by $\sim$ 3--4 MeV per nucleon. 
In symmetric nuclear matter, the BBG and the variational results are in 
agreement up to $\sim$ 1.5 $\rho_0$. Beyond this density, and in 
contrast with neutron matter, we  find 
good agreement only for the potential having spin--orbit components.

\end{abstract}


\maketitle


Homogeneous matter of nucleons plays an important role in modern nuclear 
physics. For instance, short range correlations in nuclei, induced by the 
strong nucleon--nucleon (NN) interaction, 
are expected not to be very different 
from those in infinite nuclear matter at a corresponding local density. 
This prediction has found quantitative confirmation in the studies of 
quantities like inclusive and exclusive electron scattering cross 
sections. In general, the study of 
infinite matter of nucleons, starting from a hamiltonian containing 
realistic NN interactions, shows dramatic departures from the predictions 
of the independent particle models (IPM)\cite{sick}. 
A clear example is the depletion of 
the momentum distribution, $n(k)$, at momenta below the Fermi momentum,
 $k<k_F$, and the corresponding appearence of a large momentum 
tail~\cite{fp84,gcl87,rpd89,bal90,bff90,bal91}, otherwise absent in any IPM. 
An object of intensive study in this field is the equation of state 
(EOS) of nuclear matter, both for  symmetric nuclear matter (SNM) 
and pure neutron matter (PNM). The aims are $(i)$ testing the validity of 
microscopic interactions, fitted to the properties of the light (A=2 and 3) 
nuclei, in a many--body system and $(ii)$ checking the accuracy of the adopted 
methodologies in a demanding environment. Moreover, an accurate knowledge 
of the EOS, and in particular of the density dependence of the symmetry 
energy~\cite{symmetry}, is needed in order to 
determine, with the highest possible level of confidency, 
the structure and the thermal evolution of the 
neutron stars~\cite{neutron stars}. 
In this respect, it is compulsory to use both realistic 
hamiltonians as well as reliable many--body techniques. 

From the point of view of the NN interaction, large progresses have 
been achieved in the last decade. Modern potentials~\cite{A18,nijpot,cdbon} 
are ``phase shift equivalent'', since all fit a huge set of NN scattering 
data~\cite{nijemegen} below 350 MeV with $\chi^2$ per datum close to 1.  
Three-nucleon forces have been also introduced, and the resulting 
hamiltonian provides a nice reproduction of the binding and low--lying 
states energies of light nuclei (A$\leq$10)~\cite{VMC,light}. 

In parallel to the advances in the knowledge of the nuclear interaction, 
and partly motivated by its strong state dependence, several many--body 
theories have been noticeably pushed forward. In this paper we use, and 
compare, the Correlated Basis Functions (CBF)~\cite{CBF,fafa} and the 
Bethe--Brueckner--Goldstone (BBG)~\cite{BBG} theories. 

CBF is particularly suited to deal 
with strongly interacting systems, since the non--perturbative correlation 
effects induced by the interaction are directly embedded into the basis 
states through an appropriate many--body correlation operator acting 
on some given model wave function. It is clear that complicated interactions 
are expected to induce similarly non trivial correlations. For instance, 
$(i)$ the strong one--pion--exchange potential, essential to provide the 
nuclear binding, reflects into the existence of a long--medium range 
tensor--like correlation, whereas $(ii)$ the strong, short range 
(about 0.5 fm) NN repulsion, preventing the nuclear systems from collapsing, 
generates a mostly central correlation so strong that the wave function is almost vanishing at 
these 
low internucleon distances. Operator matrix elements and expectation values 
between CBF states are, by far, more realistic than those evaluated in 
a free Fermi Gas (FG) basis. As a consequence, a CBF based perturbative 
expansion is expected to converge much faster. The obvious drawback is that 
the matrix elements are difficult to be accurately computed. The zeroth 
order of the CBF perturbative theory corresponds to the purely variational 
estimates, since the correlated ground state wave functions are fixed 
by minimizing the energy with respect to the correlation parameters. 
Within CBF the matrix elements can be computed either by cluster expansions 
in Mayer--like diagrams  
and integral equations methods or by Monte Carlo (MC) based evaluations. 
The Fermi Hypernetted Chain/Single Operator Chain (FHNC/SOC)~\cite{FHNC/SOC} 
equations belong to the first type of approach and their solution provides 
the sum of infinite classes of cluster diagrams. 
However, the FHNC--like summation 
is not complete and some diagrams, like the ``elementary'' ones, are not 
fully considered. This fact constitutes 
an approximation within the theory, that must be checked against known 
exact results, as sum rules, or compared with the 
outcomes of other methods. Variational Monte Carlo (VMC)~\cite{VMC}  
provides an alternative and exact, but expensive, way of completely sum 
the cluster contributions through stochastic evaluation of the 
needed many--body integrals. The integral equation method has the advantage 
of not being limited by the number of particles, whereas VMC becomes 
impractical at large A--values. 

Standard perturbation theory cannot be straightly applied to the 
nuclear case because of the strong, non perturbative, repulsive core.  
In the BBG theory the $in$ $medium$ 
two--body scattering $G$--matrix is introduced. The $G$--matrix has 
a regular behavior even for strong short--range repulsions, and 
it constitutes the starting point to derive the so--called {\em 
hole--line expansion}, where the perturbative diagrams are 
grouped according to the number of independent hole--lines
(see $e.g.$ Ref.\cite{BBG}). The Brueckner--Hartree--Fock (BHF) 
approximation sums only two hole--line diagrams (2HL), and it is 
expected to take accurately care of the two--body correlations. Within 
BHF, one has also to self--consistently consider the auxiliary single 
particle potential, $U(k)$, and the diagrams having $potential$ $insertions$. 
It has been recently shown that 
in the $continuous$ $choice$ for the single particle potential (see Eq.(\ref{SPP})),
the three hole--line (3HL)  
correction is  small up to densities several times larger than 
$\rho_0$~\cite{baldo-3hl}.

Quantum Monte Carlo (QMC), in its different implementations, 
as Auxiliary Field 
Diffusion MC (AFDMC) \cite{AFDMC} and Green's Function MC (GFMC)~\cite{VMC}, 
allows in principle for an $exact$ solution of the many--body Schr\"{o}dinger 
equation. However, its accuracy is limited by the $fermion$ $sign$ 
$problem$~\cite{sign-problem}, solved in the aforementioned methods by 
limiting the sample walk inside a fixed nodal surface. This $fixed$ $node$ 
approximation, which is essential to get reliable numerical estimates in 
systems with a large number of fermions, makes the present day QMC results 
essentially variational. 
The more accurate are the nodes, the closer to the true 
eigenvalues are the QMC outcomes. AFDMC and GFMC differ in the way they treat 
the spin degrees of freedom. The first method samples the spin states by  
introducing auxiliary fields of the Hubbard--Stratonovich type, whereas 
the second one sums them completely. The advantage of AFDMC 
lies in the fact that simulations with a rather large 
number of particles (up to A$\sim$ 60 for PNM~\cite{fantoni_PRL87}) 
can be carried on with a low variance. 
In contrast, GFMC has been limited so far to A=14~\cite{carlson_PRC68} in PNM, 
with the obvious consequence of showing large finite size corrections. 
None of these QMC methodologies has been so far applied to SNM.

The nuclear interaction has important tensor and 
momentum dependent (MD) components. 
The tensor potential is essentialy due to One Pion Exchange and 
is the main responsible for the nuclear binding. 
For the Argonne $v_{18}$~\cite{A18} (A18) the NN tensor 
interaction is dominant in the deuteron $(S=1,T=0)$ ground state, 
contributing  to the energy per nucleon as $<v_t>$=-8.34 MeV, whereas the 
MD potential gives only  $<v_{MD}>$=-0.52 MeV.
 However, in the $^3$H nucleus, the contributions are 
$<v_t>_{T=0}$=-10.89 MeV, $<v_{MD}>_{T=0}$=-1.97 MeV, and 
$<v_t>_{T=1}$=-0.14 MeV, $<v_{MD}>_{T=1}$=1.52 MeV~\cite{viviani_tritium}.  
We stress that the $T=1$ isospin triplet channels are the only ones 
effective in PNM. 
The evaluation in heavier systems
, either nuclei (A$>$12) or infinite matter, 
 appears methodologically problematic. 
For neutron matter, the GFMC and AFDMC estimates of the LS 
contribution employing the Argonne $v_{8'}$ (A8') ~\cite{pudliner_PRC56} 
potential are in disagreement~\cite{carlson_PRC68,fantoni_PRL87}. 
A8' is a simplified version of A18, truncated after the 
linear spin--orbit components and refitted to have the same isoscalar 
part of A18 in all the $S$ and $P$ waves, as well as in the 
$^3D_1$ wave and its coupling to the $^3S_1$. It has been used in 
GFMC calculations of light nuclei since the difference with A18 
is small and can be treated perturbatively~\cite{pudliner_PRC56}. 

In this letter we make a comparison between the BHF and CBF 
energies in  neutron and nuclear matter, paying a particular attention 
to the tensor and spin--orbit components of the NN interaction.
 The energies are computed up 
to three times $\rho_0$, considering the second order CBF perturbative 
corrections to the variational estimates and, when available, the 
three hole--line contributions to the BHF approximation. The accuracy 
of the calculations are then assessed, according to the approximations 
in the cluster and the hole--line expansions. We do not include 
three--nucleon interactions (TNI), since we are mostly interested in 
just evaluating the reliability of different many--body techniques in 
a wide range of densities. Moreover, the level of confidence in the 
available TNI is not the same as in the NN ones, both for the objective 
larger difficulties in the corresponding theoretical frameworks and 
for the much smaller three--nucleon experimental data set, to be fitted 
by the theoretical TNI.

CBF calculations in homogeneous matter of nucleons are based upon 
the set of correlated basis functions,
\begin{equation}
\Psi^{CBF}_n\, = \,( \prod_{i<j} f_{ij} )_{sym} \Psi^{FG}_n \, ,
\label{correlated}
\end{equation}
built by applying a symmetrized product of two--body correlation 
operators, $f_{ij}$, to the Fermi Gas (FG) states, $\Psi^{FG}_n$.  
In nuclear systems $f_{ij}$ depends on the internucleon distance, 
as well as on the relative state of the pair. 
We adopt a correlation operator having the same 
state dependence as the A8' potential, 
\begin{equation}
f_{ij}\, = \, \sum_{p=1,8} f^p(r_{ij})O^p_{ij} \, ,
\label{correlation}
\end{equation}
where $O^{p=1,3,5,7}_{ij}=1, \sigij , \Sij, \LS$ and  
 $O^{p=2,4,6,8}_{ij}=O^{p-1}_{ij} \tauij $, being $\Sij$ and $\LS$ 
the tensor and spin--orbit operators.

Variational estimates of the energy are obtained by minimizing the 
expectation value of the hamiltonian, 
$H=T+V=-\sum_i\hbar^2 \nabla^2_i /2m_i
+\sum_{i<j} v_{ij}$,  on the correlated ground state, 
$\Psi^{CBF}_0=( \prod_{i<j} f_{ij} )_{sym} \Psi^{FG}_0$, where 
$\Psi^{FG}_0$ is the usual FG ground state of an infinite matter of 
fermions, and we consider two--body interactions only. The correlation 
operator is determined by solving the Euler equations 
corresponding to the minimization of the energy at the two--body 
level of the cluster expansion. The free parameters of this procedure 
are the healing distances, $d_c$ and $d_t$, of the central and tensor 
components of $f_{ij}$ and the quenching factor, $\alpha$, of the spin 
dependent part of $v_{ij}$, adopted in the solution of the Euler 
equations~\cite{FHNC/SOC}. 

Within the cluster expansion method, the variational ground state energy, 
\begin{equation}
E^v_0={< \Psi^{CBF}_0|H| \Psi^{CBF}_0>}/
{< \Psi^{CBF}_0| \Psi^{CBF}_0>} \, , 
\label{var_energy}
\end{equation}
is computed by means of the FHNC/SOC theory and its 
improvements~\cite{FHNC/SOC,wiringa_PRC38}. 
The FHNC/SOC equations sum diagrams containing an infinite number 
of nucleons. 
However, the sum is incomplete since there are two 
classes of diagrams not considered: $(i)$ $elementary$ diagrams, 
contributing to the energy at least as $\rho^3$,  
since the lowest order contribution comes from four--body 
diagrams; $(ii)$ corrections, $\Delta$, 
due to the non complete inclusion of all the correlation components 
(\ref{correlation}) with $p\geq$2. 
$\Delta=0$ for state independent correlations, 
$f^{p\geq 2}=0$, whereas it is non zero in the state 
dependent case. Three--body clusters give the lowest order contribution 
to $\Delta$, which behaves at least as $\rho^2$. 
$\Delta^{MI}$ due to the momentum independent part of the correlation, 
$f^{p\leq 6}$, has been recently evaluated in SNM and 
PNM~\cite{morales_PRC66}. The results for A18,
as extracted from Tables X and XI of the Reference, 
 are:
$\Delta^{MI}$(SNM, $\rho=\rho_0$)=-2.6 MeV, 
$\Delta^{MI}$(SNM, $\rho=$1.5 $\rho_0$)=-2.8 MeV, 
$\Delta^{MI}$(PNM, $\rho=\rho_0$)=0.4 MeV, and  
$\Delta^{MI}$(PNM, $\rho=$2 $\rho_0$)=-1.4 MeV. The percentile 
contributions to the total three--body cluster energies are: $-26\%$, 
$-18\%$, $3\%$ and $-6\%$, respectively~\cite{morales_PRC66}. The magnitude of 
$\Delta^{MI}$  points to a good accuracy of the FHNC/SOC estimates in the
PNM case.
 In addition, asymmetric nuclear matter, with a low percentage of 
protons (as of interest in neutron star physics), should be safely 
addressed by the same technique.

We include in our calculations the lowerst order four--body elementary 
diagrams, $E^{(4)}_{ee}$, linear in the correlations, $[f^1(r)]^2-1$ and 
$2f^1(r)f^{p\geq 2}(r)$. $E^{(4)}_{ee}$ belongs to the 
$exchange$--$exchange$ elementary diagrams subclass 
and it is expected to be relevant for potentials having 
large Majorana components~\cite{fabrocini_PRC_Eee}. 
The other elementary diagrams have a higher power 
dependence on the correlations and are not considered in this work.

The CBF energy is evaluated by adding the second order CBF 
perturbative corrections to the variational estimate, 
$E_0^{CBF}=E_0^v+\Delta E_0^{CBF}$.  $\Delta E_0^{CBF}$ 
is computed by considering 
two--particle two--hole intermediate correlated states,
$\Psi^{CBF}_{2p2h}=( \prod_{i<j} f_{ij} )_{sym} \Psi^{FG}_{2p2h}$, 
where $\Psi^{FG}_{2p2h}$ is the FG $2p$--$2h$ state. 
$\Psi^{CBF}_{2p2h}$ is then normalized and orthogonalized to 
$\Psi^{CBF}_0$. The evaluation of $\Delta E_0^{CBF}$ is carried on 
as explained in Ref.~\cite{CBF}, 
at low (second and part of the third) orders of the cluster expansion 
of the non diagonal matrix elements of the hamiltonian. 

In the BHF approximation  the Brueckner G-Matrix, $\Gw$, is obtained 
by solving the Bethe--Goldstone equation,
\bea
&<&k_1k_2|\Gw|k_3k_4>=<k_1k_2|v|k_3k_4> \nonumber \\
&+&\sum_{k'_3k'_4} <k_1k_2|v|k'_3k'_4>
{{Q(k'_3,k'_4)}\over
{\omega-e(k'_3)-e(k'_4)}}
<k'_3k'_4|\Gw|k_3k_4> \, , 
\label{Bethe-Goldstone}
\eea
where $Q(k,k')$ is the Pauli operator ($Q(k,k')$=1 if both arguments are larger than 
$k_F$ and  $Q(k,k')$=0 otherwise) 
enforcing the scattered momenta to lie above the Fermi 
level. The single particle potential, $U(k)$, entering the definition 
of $e(k)= \hbar k^2 /2 m + U(k)$ , should be self-consistently determined
together with the G-Matrix and it is given  by  
\begin{equation}
U(k)=\sum_{k'<k_F} <kk'|G(e(k)+e(k'))|kk'>_a \, , 
\label{SPP}
\end{equation}
where the suscript $a$ indicates antisymmetrization of the matrix elements.
This choice for  $U(k)$ has been shown to considerably 
improve the convergence of the hole--line expansion~\cite{baldo-3hl}. 
Finally, the BHF energy is computed via:
\begin{equation}
E_0^{BHF}=\sum_{k_1<k_F} {{\hbar^2 k_1^2}\over {2m}} + 
{1\over 2}\sum_{k_1,k_2 < k_F} 
<k_1k_2|G(e(k_1)+e(k_2))|k_1k_2>_a \, . 
\label{E_BHF}
\end{equation}
We solve the G--matrix equation (\ref{Bethe-Goldstone}) by expanding it in 
partial waves up to $J=8$ and adopting the Born approximation for $J=9-20$. 
The accuracy of this approximation has been checked to be 
valid within a few percent \cite{baldo2001}.
The first correction to BHF arises from the  three 
hole--line diagrams. In this respect, a three--body scattering matrix, 
$T^{(3)}$, is introduced, which satisfies the Bethe--Fadeev integral 
equation~\cite{BBG,rajaraman_RMP39}. In PNM the three hole--line 
contribution is just a few percent of the BHF energy up to several 
times $\rho_0$.  
Concerning the TNI, it has  to be 
stressed that its correct evaluation  in the BBG approach 
would require their inclusion in the Bethe--Fadeev equation. However, 
TNI  are presently treated within the BHF approximation, averaging 
over the third nucleon coordinates~\cite{TNI_BHF}. In any case, as it was mentioned 
before, in the present 
paper we do not consider three-body forces. 

We first present in Table~\ref{table_A6p} the energies per nucleon 
for Argonne $v_{6'}$ (A6'), a simplified version 
of Argonne $v_{8'}$, containing only its same static part, 
without spin--orbit components 
and without refitting the phase shifts. 
 The Table gives the FHNC/SOC variational energies, $E_0^v$, 
those at the second order of the CBF perturbative expansion, $E_0^{CBF}$, 
and the BHF approximation estimates, $E_0^{BHF}$. For PNM we 
also show the three hole--line expansion~\cite{baldo04}, $E_0^{BBG}$,  
the GFMC~\cite{carlson_PRC68} and the AFDMC~\cite{sarsa03} 
 energies. 
The GFMC column gives the 
energy of 14 neutrons in a periodic box (PB), including the finite 
size box corrections, coming mostly from the long range part of 
the interaction (beyond the box  boundaries) and from the difference 
in the kinetic energies between free neutrons in a box and the 
homogeneous Fermi Gas. The box corrections for A6' were not given in 
Ref.~\cite{carlson_PRC68}, so we have used those of the A8' model, 
on the basis that the long range potential correction comes largely from 
the tensor and spin interactions, not from the spin--orbit ones.
The same number of neutrons in a periodic box has been used in the 
AFDMC estimates, but with a presumably more accurate treatment of the 
finite size corrections.
 In PNM  the 3HL diagrams  included in $E_0^{BBG}$ lower  
$E_0^{BHF}$ at any density.
The comparison with GFMC and AFDMC shows that 
in PNM the BBG and variational theories both give estimates for the energies 
accurate within a few percent. 
The CBF corrections are, by construction, 
negative and lower $E_0^v$ by $\sim$ 2 MeV at $\rho_0$. 
They rapidly increase in absolute value with 
density and this fact may be viewed as an indication that the attractive 
CBF correction might be compensated by other contributions 
(many--body cluster diagrams not included in the evaluation of the 
correlated matrix element, correlated $n>2$ particle--hole intermediate 
states and/or higher order perturbative contributions). 
The BHF and variational energies of SNM are satisfactorily close up to 
$\rho_0$, and start to differ at higher densities, 
$E_0^v$ being increasingly lower than $E_0^{BHF}$. The 
CBF corrections are as large in SNM as in PNM and  worsen the disagreement. 
It is likely that the main reason for this behavior lies in the strong 
tensor interaction. In fact, we can extract from the BHF results
$<v_t+v_{t\tau}>_{{\rm SNM}}$=-25.58 MeV 
for A6' at $\rho_0$.  In SNM there are no 
3HL/BBG and Quantum Monte Carlo results for A6'. The 3HL contributions  
in SNM have been computed for A18~\cite{baldo04} and found to be 
repulsive at low densities, up to $\sim$ $\rho_0$, and mostly attractive 
above. A quantitative evaluation of these corrections would be of great 
help to assess more completely the convergence of the hole--line 
expansion in SNM.  

Table~\ref{table_A8p} gives the energies per nucleon 
for the Argonne $v_{8'}$ interaction. Again we give 
the 3HL/BBG~\cite{baldo04}, GFMC~\cite{carlson_PRC68} and 
AFDMC~\cite{sarsa03} energies for PNM. The GFMC column reports 
in parentheses also the PNM finite size box corrections. 
These corrections are impressively large and point to the need of 
GFMC simulations with a larger number of particles in order 
to ascertain their relevance. 
We have already stressed that AFDMC seems to 
be more accurate in treating finite size effects, since the 
simulation is fully tail corrected~\cite{sarsa03}. 
The difference between box corrected GFMC and AFDMC energies at 
saturation density is small for the A6' potential, but 
it is larger in the A8' one. 
An additional source of difference may be the use of an unconstrained 
path simulation in GFMC, lowering the PNM A8' energy by $\sim$1.5 
MeV at $\rho_0$, whereas AFDMC uses a constrained path. Once more, 
the differences are smaller for A6'.  
 The 3HL diagrams in PNM lower $E_0^{BBG}$ with respect to  
  $E_0^{BHF}$ up to $\rho/\rho_0$=1.5, whereas their contribution 
is repulsive at higher densities, contrary to the A6' model. 
The CBF corrections behave as in A6', being $\sim$-2 MeV in PNM at  
$\rho_0$ and more and more attractive with increasing density. 
BBG is in good agreement with the GFMC energies, even if, 
strangely enough, BHF is in better accordance. The variational and 
CBF methods appear to work rather well up to $\rho_0$, whereas are too 
attractive beyond it. A complete analysis of the missing cluster terms 
in presence of spin--orbit terms, both in the potential and in the correlation 
(as the one performed for the spin and isospin case~\cite{morales_PRC66}), 
has not yet been done. Work in this direction is in progress. 
AFDMC is rather more repulsive than the other methods. 
Part of the discrepancy with GFMC has been resolved in Ref.~\cite{brualla03} 
by a better choice of the nodal structure of the guiding function. 
This improvement has lowered the AFDMC neutron matter energy in 
the Periodic Box  model of A8' by 1 (1.6) MeV at $\rho_0$ (2 $\rho_0$).  
 For SNM the FHNC/SOC and BHF methods give 
close energies at densities larger than $\rho_0$, whereas 
CBF appears to work better at $\rho\leq\rho_0$. 
Again we stress the need of evaluating 
the 3HL/BBG corrections. If they are $\sim$ 1--2 MeV, as for A18, then 
there are indications that for A8' FHNC/SOC is more accurate in SNM than 
in PNM and that the CBF corrections are overestimated. It results also clear 
that A8' provides a saturation point for SNM that is exceedingly too 
large and should be used with some care in studying high--density 
nuclear systems (see neutron stars). 

In Table~\ref{table_comp} the contributions to the energy per nucleon 
in BHF of the interaction components are given for the A6' and A8' models.
The BHF approach provides the total energy, but does not give access to the 
separate contributions of the potential and kinetic energy in the correlated 
ground state. However, it has been recently shown that the Hellmann-Feynmann 
theorem   can be used to calculate the separate contribution of the different 
components of the NN interaction to the potential 
energy~\cite{Hellmann-Feynmann}.  Then the kinetic energy is  evaluated 
by  substracting the potential energy from $E_{BHF}$. 
The spin--orbit potential gives a large contribution, 
$< v_{LS}(\rho=\rho_0)>\sim$ -9 MeV and 
$< v_{LS}(\rho=2\rho_0)>\sim$ -22 MeV, for both PNM and SNM. 
However, $\Delta E_{LS}=<H_{A8'}>-<H_{A6'}>$ results 
smaller, especially at $\rho_0$, where  
$\Delta E_{LS}(\rho=\rho_0)\sim$ -3.9 MeV in PNM. 
This result is in line with the other estimates, since 
$\Delta E_{LS}(\rho=\rho_0)\sim$ -3.3 MeV in BBG, 
-3.9 MeV in FHNC/SOC, -4.1 MeV in CBF, and -2.7 MeV in GFMC. 
The AFDMC energies reported here give 
$\Delta E_{LS}^{AFDMC}(\rho=\rho_0)\sim$ -.1 MeV. However, as 
pointed out before, improved estimates of  $E_0^{AFDMC}$~\cite{brualla03} 
could bring this difference down to $\sim$ -1.2 MeV.
Similar numbers are 
obtained for SNM, where $\Delta E_{LS}(\rho=\rho_0)\sim$ -4.4 MeV in BHF,
-3.1 MeV in FHNC/SOC and -3.8 MeV in CBF.

To conclude this letter, we compare in Table~\ref{table_A18} the 
SNM energies per nucleon as obtained with the A8' and A18 
potentials. The Table gives the BHF results of this paper, the 
BHF ones of Ref.~\cite{baldo2001} (BHF/BF), the BBG of the same 
paper and the FHNC/SOC. The comparison between BHF and BHF/BF is 
shown to verify how accurate are the different treatments, 
given by the two independent calculations,  of the approximations 
required  to solve the BHF equation  with a large number of partial 
waves and at high densities.   
The difference is at most of 
$\sim$ 5 $\%$ at the highest density, $\rho/\rho_0$=4.5, except where 
the A18 energies are close to zero, at $\rho/\rho_0$=3.38, 
where they are anyhow small.  
 A18 has a richer operatorial structure 
than A8', including $({\LS})^2$ and $L^2$ terms, as well as isovector 
and isotensor ones. Cluster diagrams including these extra components 
have been computed in FHNC/SOC only at some low order level (two--body 
and a few three--body diagrams). It may be expected that higher 
order diagrams become increasingly more important with the density, 
as it is suggested by the raising difference between FHNC/SOC and 
BHF/BBG at densities above $\sim$ 2 $\rho_0$.
The comparison between A18 and A8' stresses, once again, the anomalous 
behavior of the latter potential, being excessively attractive at 
large densities with respect to the complete A18 model. 
Therefore, we stress once again that one should bear in mind this fact 
if A8' is going to be used to study massive and dense objects,
 as neutron stars. 

In this paper we have compared the energies of homogeneous matter 
of nucleons (symmetric nuclear matter and pure neutron matter) as 
obtained by means of several modern many--body theories. We have 
used realistic potentials with and without spin--orbit components, 
in order to assess their importance and the accuracy of their evaluation.  
The methods employed here include  the Correlated Basis Functions and 
the Bethe--Brueckner--Goldstone theories, at different levels of 
implementation. We have also presented and discussed the quantum 
Monte Carlo results available for neutron matter. 
For neutron matter 
and in absence of spin--orbit interaction all methods, but perturbative 
CBF, are in good agreement up to $\rho\sim$ 3 $\rho_0$. 
After the inclusion of the 
LS components the agreement still persists around $\rho_0$, 
whereas it is spoiled at increasing density. In particular, the 
accuracy of the CBF approach seems to become questionable 
because of the relevance of the missing terms. All methods 
provide a contribution of the spin--orbit potential 
$\Delta E_{LS}\sim$ -3 -- -4 MeV per nucleon. The AFDMC 
and the constrained path GFMC  methods only give 
a lower contribution $\Delta E_{LS}\sim$ -0.1 -- -1.2 MeV per nucleon. 
In symmetric nuclear matter, where no Monte Carlo results are available, 
BHF and variational theories are in agreement up to $\sim$ 1.5 $\rho_0$. 
Beyond this density, we still find good agreement for the potential having 
LS components, and less without. Quantum Monte Carlo results for 
SNM are much needed to clarify the behavior of the equation of 
state at large densities.

\section*{Acknowledgments}

We acknowledge fruitful discussions with Stefano Fantoni, Marcello Baldo and 
Angels Ramos and thank Michele Viviani for providing his $^3$H results.  
This work has been partially 
supported by Grant No.  BFM2002-01868 from DGI (Spain), Grant No. 2001SGR-00064
from Generalitat de Catalunya, and 
 by the Italian MIUR through the PRIN: 
{\it Fisica Teorica del Nucleo Atomico e dei Sistemi a Molti Corpi}.  




\pagebreak

\begin{table}[h!]
\begin{center}
{
\begin{tabular}{c|cccccc|ccc|} 
& \multicolumn{6}{c|}{PNM} & \multicolumn{3}{c|}{SNM} \\ \hline  
$\rho/\rho_0$ & $E_0^v$ & $E_0^{CBF}$ & $E_0^{BHF}$ & $E_0^{BBG}$ & $E_0^{GFMC}$ & $E_0^{AFDMC}$ & $E_0^v$ & $E_0^{CBF}$ & $E_0^{BHF}$ \\ \hline
0.5 & 10.10 &  9.72 &  9.79 &  9.67 &  9.54(03)&        &  -7.54 &-9.88 &  -9.91   \\
1.0 & 15.38 & 13.61 & 15.89 & 15.09 & 14.81(11)& 15.12(4)  & -12.23 & -14.09 & -13.77   \\
1.5 & 21.17 & 18.20 & 23.02 & 21.26 & 20.65(08)&        & -15.18 & -17.49 & -14.50   \\
2.0 & 27.96 & 23.91 & 31.32 & 28.74 &      & 27.84(6)  & -16.47 & -19.43 & -13.66   \\
2.5 & 35.84 & 30.70 & 40.83 & 37.55 &      & 36.00(10)  & -16.26 & -19.62 & -11.48   \\
3.0 & 44.76 & 38.60 & 51.55 &       &      &        & -14.65 & -19.07 & -8.16   \\
\end{tabular}
}
\end{center}
\caption{Energies per nucleon, in MeV, for PNM and SNM in different 
approaches with the A6' potential. 
GFMC and AFDMC statistical errors in parenthesis. 
}
\label{table_A6p}
\end{table}

\pagebreak

\begin{table}[h!]
\begin{center}
{
\begin{tabular}{c|cccccc|ccc|} 
& \multicolumn{6}{c|}{PNM} & \multicolumn{3}{c|}{SNM} \\ \hline  
$\rho/\rho_0$ & $E_0^v$ & $E_0^{CBF}$ & $E_0^{BHF}$ & $E_0^{BBG}$ & $E_0^{GFMC}$ & $E_0^{AFDMC}$ & $E_0^v$ & $E_0^{CBF}$ & $E_0^{BHF}$  \\ \hline
0.5 &  8.74 &  8.46 &  8.28 &  8.25 &  8.40(01,-1.1) &      &  -8.99 & -11.93  & -12.02 \\
1.0 & 11.53 &  9.56 & 11.99 & 11.83 & 11.90(27,-5.1) & 14.98(6) & -15.32 & -17.86  & -18.20 \\
1.5 & 13.93 & 10.17 & 16.03 & 15.85 & 16.85(50,-11.5) &      & -20.26 & -24.06  & -22.19 \\
2.0 & 16.64 & 10.93 & 20.93 & 21.95 &      & 27.3(1) & -23.87 & -28.85 & -24.67  \\
2.5 & 19.90 & 12.80 & 26.66 & 29.04 &      & 35.3(1) & -26.22 & -32.31 & -26.03  \\
3.0 & 23.86 & 15.36 & 33.23 &       &      &          & -27.30 & -34.69 & -26.50  \\
\end{tabular}
}
\end{center}
\caption{Energies per nucleon, in MeV, for PNM and SNM in different 
approaches with the A8' potential. 
GFMC statistical errors (first number) and box corrections 
(second number) in parenthesis. 
AFDMC statistical errors in parenthesis. 
}
\label{table_A8p}
\end{table}
 
\pagebreak

\begin{table}[h!]
\begin{center}
{
\begin{tabular}{c|cc|cc|cc|cc|} 
 & \multicolumn{2}{c|}{PNM/A8'} 
 & \multicolumn{2}{c|}{PNM/A6'} 
 & \multicolumn{2}{c|}{SNM/A8'} 
 & \multicolumn{2}{c|}{SNM/A6'} 
 \\ \hline  
$\rho/\rho_0\rightarrow$ & 1.0& 2.0& 1.0& 2.0& 1.0& 2.0 & 1.0&2.0 \\ \hline 
$<v_c>$ & -6.43 & -11.08 & -7.80& -13.84 & -0.63  &  -2.62 &0.60 & -5.02   \\
$<v_\tau>$ & 1.20 & 2.79 & 1.05 & 2.58 & -1.70  & -2.36 &-3.12 & -2.59  \\
$<v_\sigma>$& -8.32 & -11.85 &-8.73 & -12.67 &-3.83  & -5.34  &-4.52 & -6.67 \\
$<v_{\sigma\tau}>$& -10.39 & -14.44 &-10.92 & -15.45 &-20.62  & -28.00 &-23.92 & -29.10 \\
$<v_t>$&0.10  & -0.14 &0.10 & -0.01 &0.27 & 0.40& 0.12 & 0.26  \\
$<v_{t\tau}>$&-3.18  &   -6.20 &-2.69 & -5.65 &-32.07  & -46.03 &-25.70 & -39.01 \\
$<v_{{\rm LS}}>$&-6.14  & -15.82 & & &-0.77  &  -7.14 & & \\
$<v_{{\rm LS}\tau}>$& -2.85 &  -7.41 & & &-8.13  &  -14.76& &  \\
\end{tabular}
}
\end{center}
\caption{
Contributions to the BHF energy per nucleon, in MeV, from the various 
components of the A6' and A8' potentials.}
\label{table_comp}
\end{table}

\pagebreak

\begin{table}[h!]
\begin{center}
{
\begin{tabular}{c|cccc|c|c} 
& \multicolumn{4}{c|}{A18} & \multicolumn{1}{c|}{A8'}& \\ \hline  
$k_F$[fm$^{-1}$] &BHF&BHF/BF&BBG&FHNC&BHF&$\rho/\rho_0$ \\ \hline
1.0 & -10.84 & -11.00 &  -7.67 &  -7.61 & -11.72 & 0.42 \\
1.2 & -13.62 & -14.13 & -12.83 &        & -15.16 & 0.73 \\
1.4 & -16.00 & -16.43 & -16.22 & -14.43 & -19.69 & 1.16 \\
1.6 & -16.12 & -16.30 & -17.00 &        & -23.49 & 1.73 \\
1.8 & -11.88 & -11.31 & -13.24 & -16.51 & -25.94 & 2.46 \\
2.0 &   0.25 &   1.14 &   1.04 &        & -26.41 & 3.38 \\
2.2 &  23.17 &  24.49 &  28.03 &  -6.88 & -24.85 & 4.50 \\
\end{tabular}
}
\end{center}
\caption{
Comparison between the BHF, the BHF/BF and BBG 
of Ref.\cite{baldo2001} and FHNC/SOC  SNM energies per 
nucleon (in MeV) for the A18. The BHF results for  A8' are also reported.
}
\label{table_A18}
\end{table}


\end{document}